\documentclass[journal,draftcls,onecolumn,12pt,twoside]{IEEEtranTCOM}
\normalsize

\usepackage{cite}

\ifCLASSINFOpdf
\else
\fi

\usepackage[cmex10]{amsmath}
\usepackage{algorithmic}

\usepackage{array}

\usepackage{mdwmath}
\usepackage{mdwtab}

\usepackage{eqparbox}

\usepackage[tight,footnotesize]{subfigure}

\usepackage{fixltx2e}

\usepackage{stfloats}

\ifCLASSOPTIONcaptionsoff
  \usepackage[nomarkers]{endfloat}
 \let\MYoriglatexcaption\caption
 \renewcommand{\caption}[2][\relax]{\MYoriglatexcaption[#2]{#2}}
\fi
 \let\MYorigsubfigure\subfigure
 \renewcommand{\subfigure}[2][\relax]{\MYorigsubfigure[]{#2}}
\usepackage{url}
\usepackage{mathtools}
\usepackage{amssymb}
\usepackage{lmodern}
\usepackage[T1]{fontenc}
\usepackage{mathrsfs}
\usepackage{subfigure} 
\usepackage{url}
\usepackage{bbm}
\usepackage{xcolor}
\usepackage{soul}
\usepackage{setspace}
\usepackage{float}
\usepackage{soul}
\usepackage{xcolor}

\makeatletter
\newcommand{\@giventhatstar}[2]{#1\,\middle|\,#2}
\newcommand{\@giventhatnostar}[3][]{#1(#2\,#1|\,#3#1)}
\newcommand{\giventhat}{\@ifstar\@giventhatstar\@giventhatnostar}
\makeatother

\begin{document}
%
\title{Free-Space Optical Communications with Detector Arrays}

\author{Muhammad~Salman~Bashir,~\IEEEmembership{Member,~IEEE}
\thanks{
}
\thanks{}
\thanks{} 
}

%
%

\markboth{}%
{}
%



\maketitle

\begin{abstract}
Detector arrays are commonly used for free-space optical communications in deep space. Such detector arrays---by virtue of their size---help in the collection of the optical signal even when there is some misalignment between the transmitter and receiver systems. In this paper, we argue that for the common Gaussian beam profile, a detector array receiver is more useful for minimizing the probability of error than a single detector receiver of the same dimensions. Furthermore, the improvement in the error probability is more pronounced for low signal-to-noise ratio conditions, and the probability of error decreases monotonically as a function of the number of detectors in the array. However, communication with detector arrays results in a larger computational complexity at the receiver. Additionally, such detector arrays are also more advantageous for beam position tracking on the detector array in order to minimize the pointing loss. 
\end{abstract}


%
\IEEEpeerreviewmaketitle

\section{Introduction}
\begin{figure}
	\subfigure[]{\includegraphics[scale=0.8]{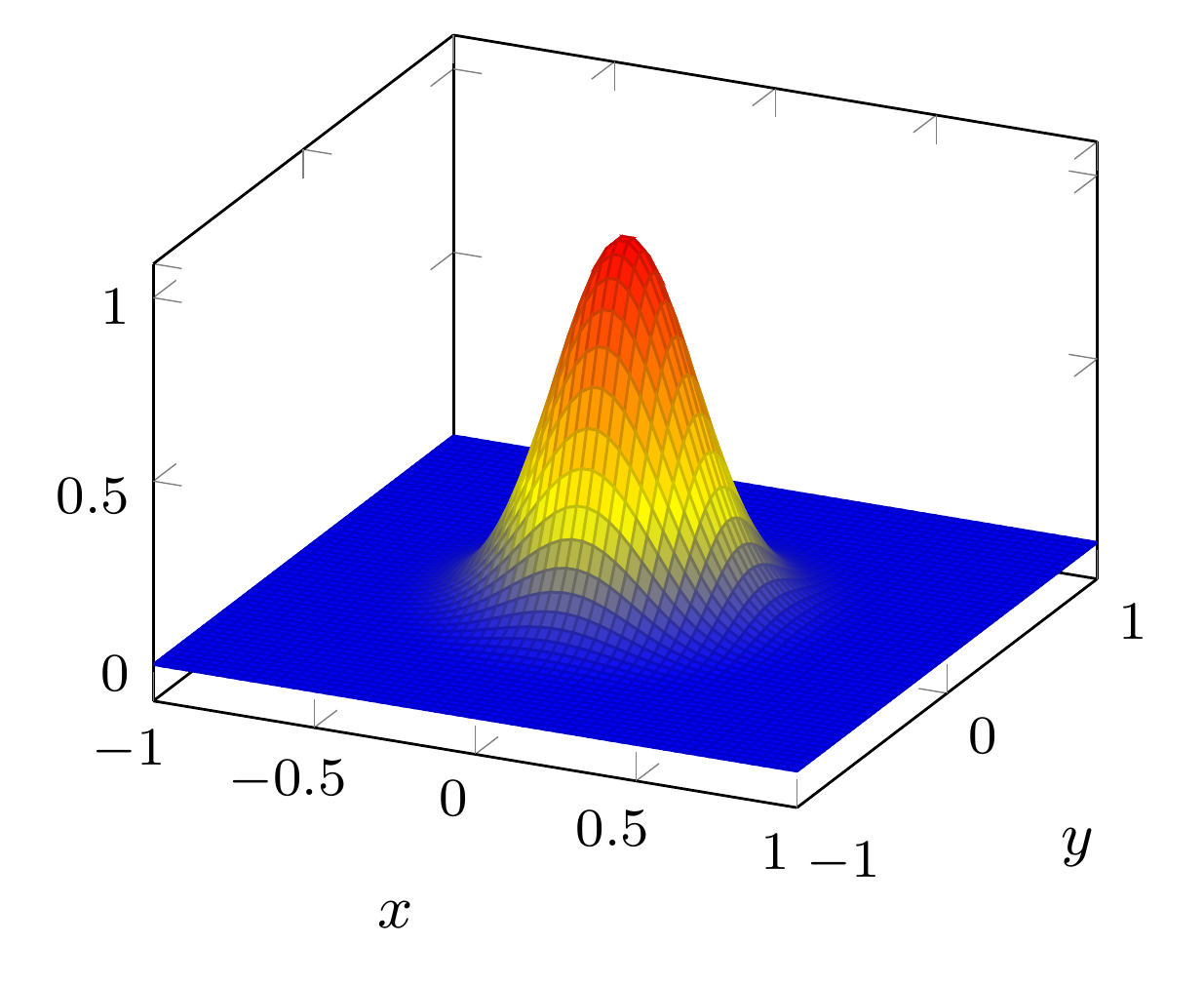}}
	\subfigure[]{\includegraphics[scale=0.9]{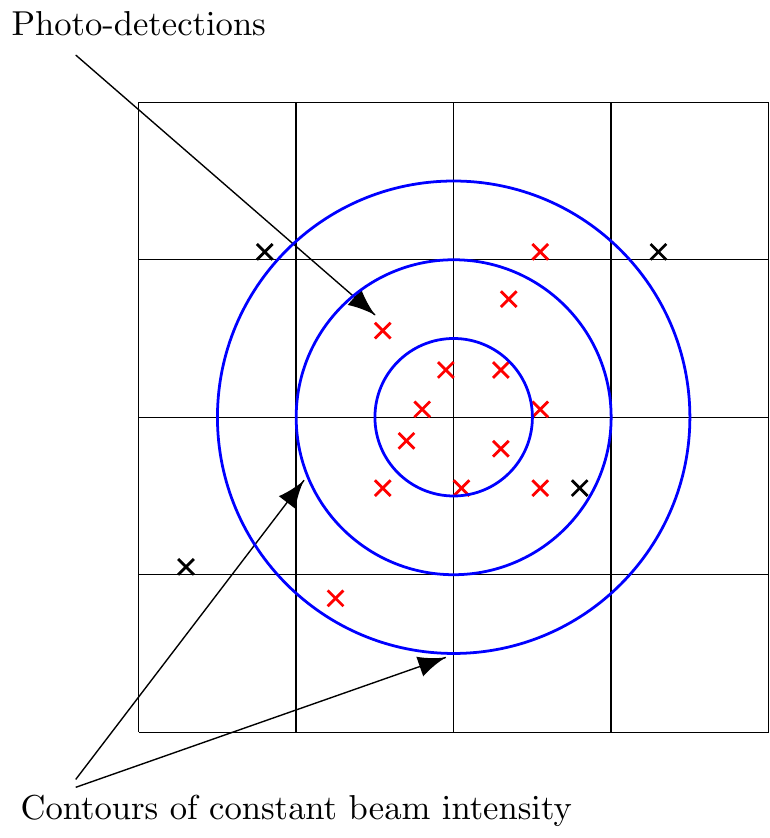}}
	\caption{The Gaussian intensity profile on a detector array for an FSO channel and the resulting photodetections observed on a $4\times 4$ detector array during a given observation period. The signal and noise photodetections are depicted by red and black crosses, respectively.} \label{gauss_beam}
\end{figure}
Free-space optics (FSO) is a promising wireless communications technique that supports high data-rate communications. Such FSO communication also supports large link distances due to the smaller divergence angle of the transmitted optical signal as compared to the radio frequency systems. The smaller divergence angel results from the smaller wavelength of the optical frequencies $(\theta_{\text{div}}\approxeq \lambda/D$, where $\lambda$ is wavelength and $D$ is the transmit telescope aperture), and necessitates the employment of pointing and tracking techniques  for the purpose of alignment between the transmitter and receiver \cite{Kaymak}. An FSO system typically consists of a laser transmitter that transmits a narrow beam of light which is sensed/detected by a photosensitive detector or an array of detectors at the receiver. 
\begin{figure}
	\centering
	\includegraphics[scale=1.3]{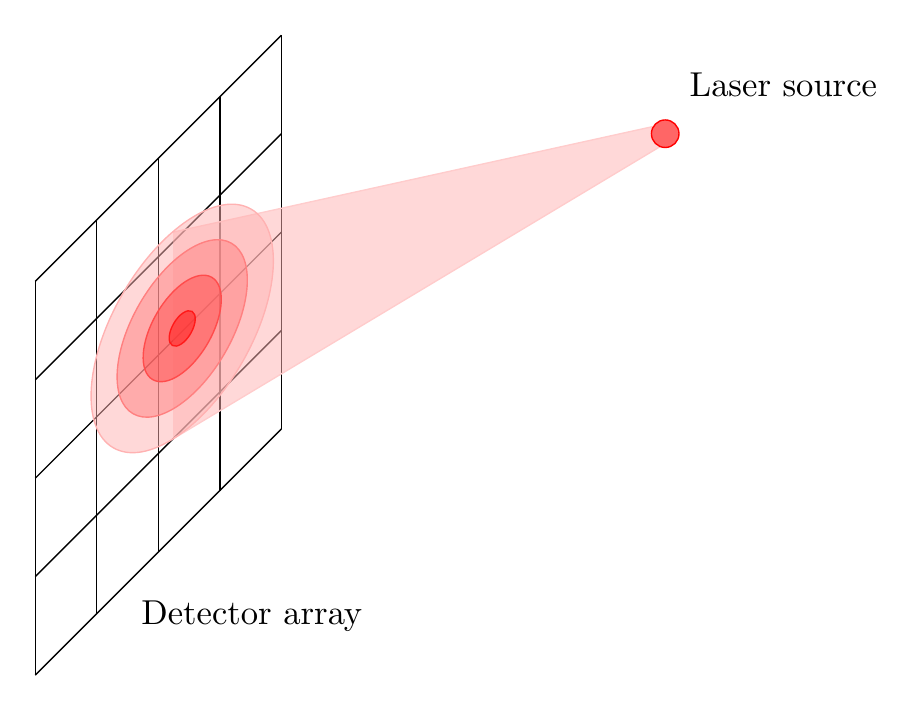}
	\caption{Gaussian beam contours on a detector array of 16 cells.}
	\label{detector_array}
\end{figure}


\subsection{Optical communication with detector arrays}\label{optic_comm}

Detector arrays---or \emph{focal plane arrays} (FPA)---have commonly been used for deep space optical communications as discussed by authors in \cite{Srinivasan1},    \cite{Vilnrotter}  \cite{Srinivasan} and \cite{Vilnrotter1}.  For an ideal turbulence free channel that spans a long distance, the optical field on the aperture plane can be approximated by a plane wave that has a uniform value over the plane \cite{Kaushal}.  By the Fourier optics theory, the optical field at the FPA and the optical field at the aperture planes form the Fourier Transform pairs\footnote{Strictly, the field at the focal lengths before and after the lens form Fourier Transform pairs.} \cite{Goodman}. Hence, for a turbulence free channel, the point-spread function on the FPA resembles a narrow function similar to the Dirac Delta function. However, due to turbulence in a practical system, the optical field is not uniformly distributed on the aperture plane, and this leads to the distortion or spreading of the point-spread function on the FPA. Moreover, the centroid of the received optical beam can also wander about its mean position. This warrants the use of detector arrays at the receiver which increase the collection area in order to be able to receive most of the signal energy at the receiver end.  Even with adaptive optics techniques, the point-spread function may be large enough to justify the use of detector arrays \cite{Srinivasan2}.

A commonly used model of the intensity distribution on the FPA (for a turbulence free channel) is the Gaussian function which results from the Gaussian beam impinging on the aperture plane\footnote{This is true because the Fourier transform of a Gaussian function is another Gaussian function.} \cite{Najafi}, \cite{Bashir1} and \cite{SnyderandMiller}. Fig.~\ref{gauss_beam} shows such a Gaussian profile. It is not uncommon for the beam width factor $\rho$ to be a fraction of a meter (about 0.25 meters) in length under usual normal weather conditions and a link distance of about 1 kilometer \cite{Najafi}. Hence, it is possible that the beam will illuminate a sufficiently large number of detectors of the array if a suitable (small) size for such detectors is chosen. A commonly used optical receiver is the \emph{intensity modulated/direct detection} (IM/DD) receiver that detects the energy of the received signal by counting the number of photodetections during some observation interval, and the photon counts are then used to make a symbol decision \cite{Bashir3}. Such photon counting receivers are commonly used in deep space communications \cite{Srinivasan1}. The number and locations of these photodetections are governed by a \emph{Gaussian Poisson point process} \cite{SnyderandMiller}. Most commonly used modulation schemes for IM/DD systems are \emph{pulse position modulation} (PPM) and \emph{on-off keying} (OOK).

The back ground radiation and the thermal effect of the detector array acts as a source of noise in an optical communication system. The intensity of the background radiation is modeled as a constant function on the surface of the detector array. In other words, the background radiation illuminates the detector array uniformly. The resulting photodetections due to background radiation as well as thermal noise are modeled by a \emph{homogeneous Poisson point process}.

The number of photodetections (signal plus noise) reported by each detector during an observation interval forms a \emph{sufficient statistic} for detecting the received optical symbol. 

\section{Literature review and contributions of this paper}

 Ideally, if the intensity function can be characterized analytically, an optimal weighting scheme can be devised that will weight the photon count in each detector according to the signal-to-noise ratio (SNR) in each detector before combining them. However, due to turbulence, it may be difficult to characterize the received signal intensity and suboptimum weighting techniques will have to be considered. Vilnrotter  \emph{et. al} in \cite{Vilnrotter} and \cite{Vilnrotter1} have proposed a suboptimum weighting technique to weight the photon counts in each detector before they are combined and processed by the digital signal processing blocks. They have used a binary weighting scheme in which the photon counts from the detectors exposed to the largest values of the signal intensity (or the largest signal-to-noise ratio) are assigned a value ``1,'' whereas the rest of the detector counts are assigned a value ``0.'' Hence, only the detectors with the best signal-to-noise ratio are incorporated in the signal detection process. 

However, for many FSO channels such as the stratosphere that establishes the balloon-to-balloon and the balloon-to-satellite links for the Google's project Loon, the atmospheric turbulence affects are negligible. For such channels, the intensity function on the FPA can be characterized analytically, and depends on the transmitted beam profile. For such channels, we propose a general optimal weighting scheme for the hard decision channel (optimum in the sense of minimization of the error probability)  that is based on the work in \cite{Bashir3}.  

In this paper, we make the case for using an array of smaller detectors at the optical  receiver instead of using one large detector of the same dimensions. Our claim is that the probability of error for signal detection with an array of detectors is smaller than a single detector, and the improvement in the probability of error is substantial at low signal-to-noise ratio. Even though the Peyronel \emph{et. al} have proposed large sized \emph{luminescent detectors} for optical signal detection at high frequencies, most semiconductor diodes in use today incur a significant increase in response times (and hence lower bandwidth) if the area is increased beyond a certain limit \cite{Peyronel}. Additionally, an array of detectors is more robust than using a single detector if the system breaks down due to a fault in a given detector. Hence, the use of an array of smaller detectors in this regard is a much better choice than a single detector from the perspectives of probability of error, response time and fault tolerance. Having said that, in order to minimize the error probability with detector arrays, we need to know the intensity profile of the received signal on the FPA, which in many cases is known or approximately known \cite{Kaushal}.

The paper is organized as follows. In the next section, we will use an intuitive argument to show that  for a fixed FPA area, the probability of error can only decrease monotonically if the number of photodetectors is increased (or the size of each cell or photodetector becomes smaller). We will also support this argument with a mathematical formulation.  We will also discuss about the ``continuous'' array which, performance-wise,  bounds the probability of error for any detector array from below. 

Furthermore, in the later sections we will observe the probability of error performance of detector arrays for different scenarios of signal-to-noise ratio where we will see that the performance gets better with the number of detectors $M$. However, the better performance of detector arrays with large $M$ comes at a cost; therefore,  a complexity analysis of detector arrays is also presented. The overall results of this paper are summed up in Section~\ref{conclude}. The appendix contains the derivation of the probability of error for continuous arrays. 


\section{Probability of error with a detector array}\label{Pe}

Let the intensity of the incident optical signal be a general function $\lambda_s(x,y)$ on the two dimensional focal plane array. Moreover, let us assume that the background radiation and thermal effect of the detector array is modeled by a constant intensity $\lambda_n$. Let us further assume that the FPA has $M$ square shaped detectors of uniform area, and the photodetection count $Z_m$ in the $m$th cell or detector of the detector array is modeled by a {nonhomogeneous Poisson point process} during a given observation interval \cite{Bashir1}:
\begin{align}
P(\{ Z_m = z_m \}) = \frac{ \exp \left( {-\iint_{A_m} \left[ \lambda_s(x,y) + \lambda_n \right] \, dx\, dy} \right) (\iint_{A_m} \left[ \lambda_s(x,y) + \lambda_n \right] \, dx\, dy)^{z_m} } {z_m!} ,
\end{align}
where $A_m$ is the region of the $m$th detector on the FPA and $Z_1, Z_2, \dots, Z_M$ are independent Poisson random variables.

For the PPM scheme, a hard-decision maximum likelihood receiver that operates on a symbol-by-symbol\footnote{For the sake of simplicity, we consider an uncoded system that does not utilize any error correcting codes.} basis will differentiate between the following two hypotheses during the observation interval: $H_0$ that no signal beam exists in a given slot of the PPM symbol, and $H_1$ that there is a signal beam in the given slot. The observation interval is the slot width in this case. For such a receiver and a $K$ symbol PPM scheme, it can be shown that the probability of a correct decision given that symbol $j$ $(1 \leq j \leq K)$ is transmitted is \cite{Bashir3}
\begin{align}
P_{c|j} = \left[P\left( \left\{  \sum_{m=1}^M (Z_m^{(j)}  - Z_m^{(i)}) \alpha_m >0  \right\} \right) \right]^{K-1}  \quad  i \neq j, \label{alphas}
\end{align}
where $Z_m^{(j)}$ and $Z_m^{(i)}$ represent the random photon counts in the $j$th and $i$th slots of the PPM symbol, respectively ($Z_m^{(j)}$ corresponds to the ``signal+noise'' photons, and $Z_m^{(i)}$ corresponds to the "noise only" photons), and $i$ represents any ``noise only'' slot. For the equiprobable symbol scheme, $P_c = P_{c|j}$ and the probability of error is just 
\begin{align} \label{Pe3}
P_e = 1 - P_c = 1 - \left[P\left( \left\{  \sum_{m=1}^M (Z_m^{(j)}  - Z_m^{(i)}) \alpha_m >0  \right\} \right) \right]^{K-1}.
\end{align} The weights $\alpha_m$ are defined as \cite{Bashir3}
\begin{align}
\alpha_m =  \ln \left(  1+ \frac{1}{\lambda_n A^{(m)}} \iint\limits_{A_m} \lambda_s(x,y) \, dx\, dy \right), \label{alphas1}
\end{align}
where $A^{(M)}$ is the (uniform) area of the region $A_m$ for an array of $M$ detectors. The signal power in the $m$th cell is $\iint\limits_{A_m} \lambda_s(x,y) \, dx\, dy$ and the noise power is $\lambda_n A^{(M)}.$ Alternatively, \eqref{alphas1} can be defined as 
\begin{align}
	\alpha_m = \ln(1+\text{SNR}_m),
		\end{align}
where $\text{SNR}_m$ is the signal-to-noise ratio in the $m$th detector of the array. Since the maximum likelihood detector minimizes the probability of error\footnote{Strictly speaking, the maximum a posteriori probability (MAP) detector minimizes the probability of error. However, for the common case of equiprobable symbols, the MAP and ML detectors are exactly the same.}, the weights $\alpha_m$ defined in \eqref{alphas1} are optimum.

  In order to maximize the probability in \eqref{alphas} or minimize the probability of error, we need to maximize $P \left( \left\{ \sum_{m=1}^{M} ( Z_m^{(j)} - Z_m^{(i)} ) \alpha_m >0  \right\} \right).$ However, it is not straightforward to compute this probability measure analytically. Nonetheless, we will use a heuristic argument to argue that the probability of the correct decision increases monotonically as we make the array finer, or, in other words, increase $M.$ Intuitively, in order to increase $P \left( \left\{ \sum_{m=1}^{M} ( Z_m^{(j)} - Z_m^{(i)} ) \alpha_m >0  \right\} \right)$, we need to weight the signal photons as large as possible compared to the noise photons. We also know that the closer a given cell is to the peak of $\lambda_s(x,y),$ the larger the signal photons in that cell. If the partition is not fine enough, all the signal and the noise photons in a large cell are weighted by the same factor $\alpha_m.$ However, breaking this large cell into smaller subcells results in a better weighting scheme since most of the signal photons will now have a larger weight on average than the noise photons in the said subcells. 

Alternatively, $Y_m \triangleq Z_m^{(j)}-Z_m^{(i)}$ for the $m$th cell represents a \emph{Skellam} random variable with mean $\mu_m=\iint_{A_m} \lambda_s(x,y)\, dx \, dy, $ and variance $\sigma^2_m =\iint_{A_m} \lambda_s(x,y)\, dx \, dy + 2 \lambda_n A$  \cite{Bashir3}. Let us now define 
\begin{align}
X \triangleq \sum_{m=1}^M \alpha_m Y_m,
\end{align}
and we need to maximize $P(\left\{X > 0 \right\})$ as a function of $M$ in order to minimize the probability of error. In the remaining section, we will consider the minimization argument for the Gaussian beam, which is a common intensity profile for free-space optical communications as discussed in Section~\ref{optic_comm}.

Let the detector plane be perpendicular to the $z$ axis and located at a distance $z$ meters from the transmitter. For the common case of a Gaussian beam, the intensity at a point $(x,y)$ on the detector array is given by \cite{Najafi}
\begin{align}
\lambda_s(x,y,z) = \frac{I_0}{\rho^2(z)} \exp\left( \frac{-(x-x_0)^2 -(y-y_0)^2}{2 \rho^2(z)}  \right),
\end{align}
where $\rho(z) = \rho_0 \sqrt{1 + \left(\frac{\lambda z}{\pi \rho_0^2} \right)^2}$ meters, and $(x_0, y_0)$ is the center of the Gaussian beam on the detector array. The factor $\rho_0$ is the \emph{beam waist}, and $\rho(z)$ is known at the \emph{beam radius} or the \emph{spot size}. The factor $I_0$ is a constant measured in Watts/$\text{meters}^2$/seconds, and the peak intensity at a distance of $z$ meters from the transmitter is $I_0/\rho^2(z).$

For $M$ large enough, $X$ is approximately distributed as a Gaussian by the \emph{Lindeberg-Feller central limit theorem} \cite{Hajek}. The mean and standard deviation are given by
 \begin{align}
\mu_X&= \sum_{m=0}^{M} \ln(1+\text{SNR}_m) \iint_{A_m} \frac{I_0}{\rho^2(z)} e^{-\frac{(x-x_0)^2 + (y-y_0)^2}{2 \rho^2(z)}} \, dx \, dy \label{mu}\\
\sigma_X &= \sqrt{\sum_{m=0}^{M}\left[\ln(1+\text{SNR}_m)\right]^2 \left[\iint_{A_m} \frac{I_0}{\rho^2(z)} e^{-\frac{(x-x_0)^2 + (y-y_0)^2}{2 \rho^2(z)}}\, dx \, dy + 2 \lambda_n A^{(M)}  \right] } \label{sigma},
 \end{align}
and $\displaystyle P(\{ X > 0 \})  \approxeq Q\left(  \frac{-\mu_X}{\sigma_X}  \right) $ because of the Gaussian approximation of $X$. For the special case of the poor signal-to-noise ratio, we can show that the probability of error decreases monotonically with $M.$ For small real number $\nu$ such that $|\nu| < 1$, it can be shown that \cite{Gallager},
\begin{align} \label{expansion}
\ln(1 + \nu) = \nu  - \frac{\nu^2}{2} + \frac{\nu^3}{3} - \frac{\nu^4}{4} + \frac{\nu^5}{5} - \cdots
\end{align}
For the sake of brevity, let the signal power in the $m$th cell be denoted by $I_0 s_m.$ Then
\begin{align}
s_m = \iint_{A_m}\frac{1}{\rho^2(z)} e^{-\frac{(x-x_0)^2 + (y-y_0)^2}{2 \rho^2(z)}} \, dx \, dy,
\end{align}
and $\mu_X$ and $\sigma_X$ in \eqref{mu} and \eqref{sigma}, respectively, can be rewritten as 
\begin{align}
\mu_X&= \sum_{m=1}^{M} \ln\left(1+ \frac{I_0 s_m}{\lambda_n A^{(M)}} \right) I_0 s_m, \label{mu1}\\
\sigma_X &= \sqrt{\sum_{m=1}^{M}\left[\ln \left(1+\frac{I_0 s_m}{\lambda_n A^{(M)}} \right) \right]^2 (I_0 s_m + 2 \lambda_n A^{(M)}) } \label{sigma1}.
\end{align}
For the case of poor signal-to-noise ratio, $\displaystyle \frac{I_0 s_m}{\lambda_n A^{(M)}} << 1$ for all $m$. Hence, using the result of the Taylor series expansion in \eqref{expansion}, and ignoring the second and higher order terms, we have that 
\begin{align}
\mu_X &\approxeq  \sum_{m=1}^M \frac{I_0^2 s_m^2}{\lambda_n A^{(M)}}, \\
\sigma_X &\approxeq  \sqrt{\sum_{m=1}^M \left(\frac{I_0 s_m}{\lambda_n A^{(M)}}\right)^2 (I_0 s_m + 2 \lambda_n A^{(M)})}\nonumber \\
&= \sqrt{ \sum_{m=1}^M \left[ \frac{(I_0 s_m)^2}{(\lambda_n A^{(M)})^2}    + 2 \frac{I_0 s_m}{\lambda_n A^{(M)}} \right] I_0 s_m }, \label{snr}
\end{align}
and since the signal-to-noise ratio is much smaller than 1, the second order term in \eqref{snr} may be ignored so that
\begin{align}
\sigma_X \approxeq \sqrt{ \sum_{m=1}^M 2 \frac{(I_0 s_m)^2}{\lambda_n A^{(M)}} }.
\end{align}
The probability of error is
\begin{align} \label{Pe1}
P_e \approxeq 1 - \left[ Q\left( - \frac{I_0}{\sqrt{2 \lambda_n}} \sqrt{\sum_{m=1}^M \frac{s_m^2}{A^{(M)}}} \right) \right]^{K-1}. 
\end{align}
\begin{figure}
	\centering
	\includegraphics[scale=1.5]{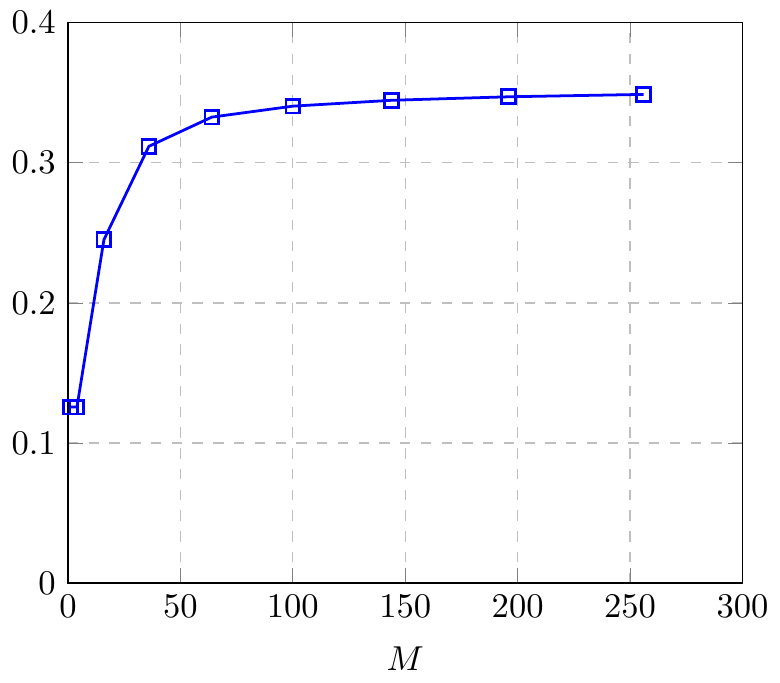}
	\caption{Plot of the quantity $\displaystyle \sqrt{\sum_{m=1}^M \frac{s_m^2}{A^{(M)}}}$ as a function of $M$ for a Gaussian beam.} \label{fig1}
\end{figure}
We plot the function $ \displaystyle \sqrt{ \sum_{m=1}^M \frac{s_m^2}{A^{(M)}} }$ as a function of $M$ in Fig.~\ref{fig1}. For this plot, we use a square detector array (with each detector also being of a square shape with uniform area) of area 4 $\text{meters}^2$. The detector array extends from -1 to 1 meters along each of the $x$ and $y$ axis. The value of $\rho(z)$ was set at 0.2 meters, and the beam center position $(x_0, y_0)$ was fixed at $(0, 0).$ 

 We note that the value of the function $ \displaystyle \sqrt{ \sum_{m=1}^M \frac{s_m^2}{A^{(M)}} }$ is the same for $M=1$ and $M=4$. However, it increases monotonically with $M$ for larger values of $M$ such as 16, 36, 64 etc. The fact that the function has the same value for $M=1$ and $M=4$ agrees with our earlier intuition that the photons near the center of the beam are more likely to be signal photons than the noise photons, and therefore, ought to be weighted more heavily than the photons that are far away. However, due to the circularly symmetric nature of the Gaussian beam and the fact that it is centered at $(0,0)$, the photons near the center of the beam have the same weight as the photons that are further away for both $M=1$ and $M=4$ scenarios. Hence, the probability of error is the same for both the cases. However, for larger values of $M$, the weighting factor of the photons near the beam center improves, which results in a decrease in the error probability.
 
   Since $ \sqrt{ \sum_{m=1}^M \frac{s_m^2}{A^{(M)}} }$ increases monotonically with $M$, the quantity $-\frac{I_0}{\sqrt{2\lambda_n}} \sqrt{ \sum_{m=1}^M \frac{s_m^2}{A^{(M)}} }$ decreases montonically with $M$ since both $I_0$ and $\lambda_n$ are positive. Since the $Q$ function is a strictly decreasing function, we have that the $Q(\cdot)$ will increase monotonically with $M$, which implies by \eqref{Pe1} that the probability of error will decrease as $M$ becomes large.
   \subsection{Lower bounding the probability of error with continuous arrays}
   In order to lower bound the probability of error with detector arrays, we introduce the notion of a \emph{continuous array} \cite{Bashir1}. A continuous array is a limiting form of a \emph{discrete array} where the discrete arrays are the practical detector arrays that have been discussed earlier in this paper. Let us define a continuous array more formally as follows. 
   \paragraph*{Definition:} For a fixed array area $A_d$,
   \begin{center}
   	discrete array $\to$ continuous array as $M \to \infty.$
   \end{center}
In other words, as $M \to \infty,$ $A^{(M)} \to 0$ for a continuous array. This mean that we will have the exact knowledge about the location of a given photodetection on a continuous array as opposed to a discrete array where the location of a photodetection is only known to exist within a cell. We can also say that the exact location of a photodetection is quantized to the center of the cell or detector in which it occurs.

Since we have access to the exact location information of a photodetection in case of a continuous array, we can weight it optimally for the purpose of symbol detection. Therefore, by virtue of the heuristic argument laid down in the third paragraph of Section~\ref{Pe}, a continuous array will achieve the minimum probability of error of all the discrete arrays. The probability of a correct decision for a continuous array has been derived in detail in  the appendix for a hard-decision maximum likelihood receiver that detects the PPM symbols of order $K$ for an uncoded system. For an equiprobable symbol transmission, the probability of error is decision given a symbol
\begin{small}
	\begin{align}
	P_{e} &= 1 - \left( \left\{  (n_j - n_k) \ln \left(\int_{-1}^1 \int_{-1}^{1} \left(\frac{I_0}{ \rho^2(z)} e^{ \frac{-(x-x_0)^2 - (y-y_0)^2}{2\rho^2(z)} } + \lambda_n\right) \, dx\, dy \right)+ \sum_{i=1}^{n_j} \ln \left( \frac{I_0}{ \rho^2(z)} e^{ \frac{-(x^{(j)}_i-x_0)^2 - (y^{(j)}_i-y_0)^2}{2\rho^2(z)} } + \lambda_n \right) \right. \right.  \nonumber \\
	& \left. \left. -  \sum_{i=1}^{n_k} \ln \left( \frac{I_0}{ \rho^2(z)} e^{ \frac{-(x^{(k)}_i-x_0)^2 - (y^{(k)}_i-y_0)^2}{2\rho^2(z)} } + \lambda_n \right)   - 2(n_j - n_k) \ln (\lambda_n A_d)      > 0  \right\} \right)^{K-1},
	\end{align}
\end{small}
where $n_j$ and $n_k$ are the photon counts in the ``signal plus noise'' and ``noise only'' slots, respectively. Moreover, $(x_i^{(j)}, y_i^{(j)})$ and $(x_i^{(k)}, y_i^{(k)})$ are the exact locations of the $i$th photodetection in the $j$th and $k$th slots of the PPM symbol, respectively, and $A_d$ is the total area of the detector array. For the sake of clarity, we note that the $k$th slot corresponds to any noise only slot.

\section{Simulations}
Let us consider \eqref{alphas} again where the probability of a correct decision for the maximum likelihood receiver depends not only on the photon counts $Z_m$ in the $m$th detector of the array, but also on the weighting factors $\alpha_m$. For a Gaussian beam, these photon counts and the weights are functions of the parameters such as peak intensity $I_0$, the beam radius $\rho$, the noise intensity $\lambda_n$ and the beam position on the detector array $(x_0, y_0)$. The knowledge of these, possibly unknown, parameters is required in order to i) weight the photon counts accurately for optimal detection \cite{Bashir3} and ii) specifically, the beam position $(x_0, y_0)$ information is required to align the transmitter and receiver assemblies.  In contrast, a single detector receiver $(M=1)$ does not require the knowledge of these parameters for optimal detection.

For short distance (1 kilometer or less) and relatively turbulent free channels in the stratosphere such as those associated with Google's project Loon, the channel conditions are not likely to be adverse. For such channels, the Gaussian profile of the intensity will likely hold at the detector array, and the peak intensity and beam radius may also be known, although the beam center position on the array may still be time-varying and unknown due to air currents perturbing the balloons about their mean positions. However, for longer distances in a free-space optics channel, the random atmospheric effects such as turbulence and scattering may become more significant so as to cause a change in the beam shape and other related parameters at the receiver \cite{Kaushal}.

In the remaining sections, we consider the effect of number of detectors in the detector array on the error probability for a system that employs Gaussian beams. We consider different scenarios of signal-to-noise ratio in this regard. Furthermore, we compare the error probabilities for an ideal system that possesses perfect knowledge of the Gaussian beam parameters, and a practical system where there is uncertainty about the parameters.

For the purpose of computing the probabilities of error, let us assume that we have a square detector array where each detector is also a square of uniform area, and the total area of the array is 4 $\text{meters}^2.$ Moreover, the center of the detector array lies at $(0, 0).$ Let us further assume that the received laser power $P_R$ is $1\times10^{-5}$ Watts. For a 30 Gbps 8-PPM scheme, the slot width $T_s$ of the PPM symbol turns out to be about $1\times 10^{-11}$ seconds. Let us assume that the wavelength $\lambda$ of the signal is 1550 nanometers. Then the energy in each photon is $hc/\lambda$, where $h$ is the \emph{Planck's constant} and its value is $6.62607004 \times 10^{-34} {m}^2 kg / s$, and $c$ is the speed of light in vacuum which is $3\times 10^{-8}  m/s$. Let the photoconversion efficiency of the photodetector be $\eta=0.5.$ Furthermore, let us denote the average number of signal photons in a PPM slot by $n_s$ and the noise photons by $n_b.$ Hence, on the average, there are \[ n_s =  \frac{P_R T_s \eta}{hc/\lambda} =  \frac{1 \times 10^{-5} \times 1 \times 10^{-11} \times 0.5} {1.28 \times 10^{-19} } \approxeq 400 \text{ photons}.  \]
Thus, we are looking at a total of few hundred signal photons detected on average by the entire FPA for the said received signal power.

Since the probability of error for detector arrays also depends on the beam center position $(x_0, y_0)$, we compute the probability of error performance for two scenarios: i) when the beam position is fixed on the array, (ii) and when it is varied in order to get an idea of the ``average'' probability of error (where the average is computed with respect to the beam center position).  In this regard, we sample the beam position coordinates $x_0$ and $y_0$  independently and uniformly between -0.75 and 0.75, and probability of error is computed for each realization $(x_0, y_0).$ In the end, the average probability of error is computed.

Since there is no closed form expression for computing \eqref{Pe3}, we used Monte Carlo based simulations for computing the probabilities of error for $M = 4$ and $M=16$ cases.  For $M \geq 64,$ we used the approximate closed form expression for the probability of error given by \eqref{Pe1}. For $M=1$, the probability of error for $K$-PPM is 
\begin{align}
P_e = 1 - \left( \sum_{k=1}^\infty P(\{  X = k \})   \right)^{K-1},
\end{align} 
where $X$ is a Skellam random variable with mean $\iint_C \frac{I_0}{\rho^2(z)} e^{-\frac{(x-x_0)^2 + (y-y_0)^2}{2 \rho^2(z)}} \, dx \, dy $ and variance $\iint_ C\frac{I_0}{\rho^2(z)} e^{-\frac{(x-x_0)^2 + (y-y_0)^2}{2 \rho^2(z)}} \, dx \, dy + 2 \lambda_n A_d$, where $C$ is the region of the detector array.
\subsection{Probability of error with perfect knowledge of parameters} \label{certain}

\begin{figure}
	\centering
	\includegraphics[scale=1.5]{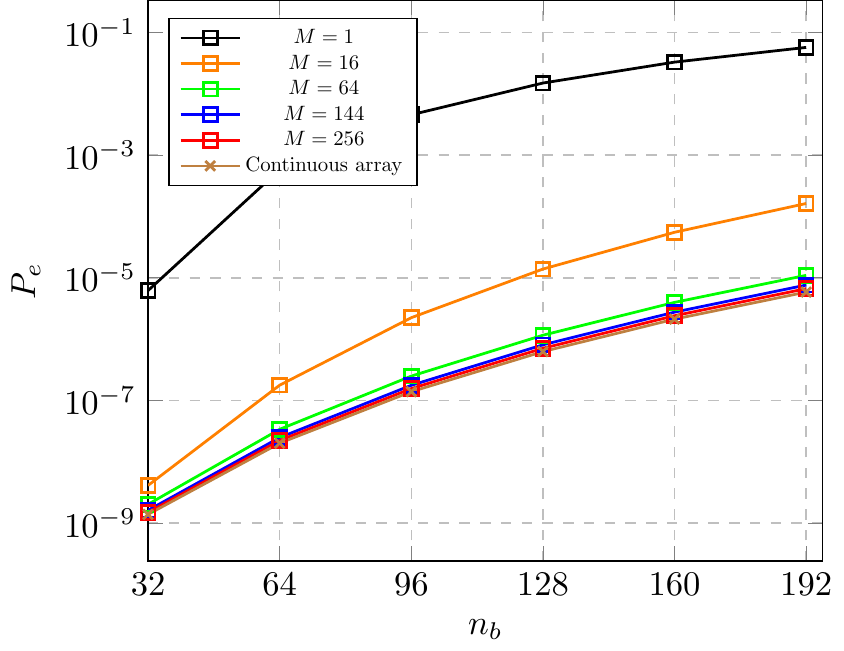}
	\caption{Probability of symbol error as a function of average number of noise photons for $I_0/\rho^2(z)=200$, $\rho(z) = 0.2$ and $(x_0, y_0) = (0,0).$ } \label{fig2}
\end{figure}

In this section, we assume that the parameters of the Gaussian beam are perfectly known at the receiver. 

\begin{figure}
	\centering
	\includegraphics[scale=1.5]{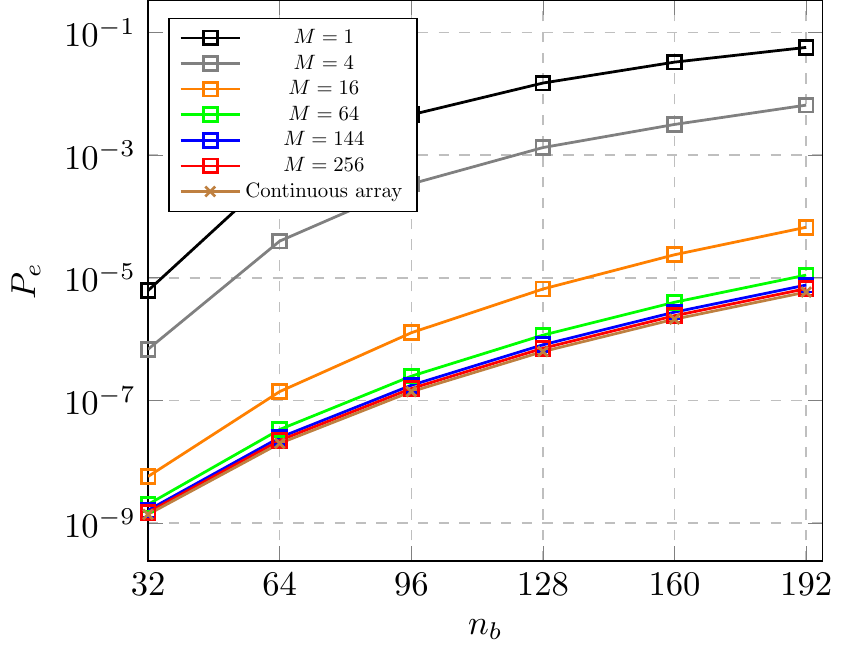}
	\caption{Probability of symbol error as a function of average number of noise photons for $I_0/\rho^2(z)=200$, $\rho(z) = 0.2$ and $(x_0, y_0)$ randomly sampled. } \label{fig4}
\end{figure}

Fig.~\ref{fig2} shows the probability of error performance as a function of average noise photons $n_b.$ For these curves, the average signal power is constant with $n_s = 50$ photons and $\rho(z)$ equal to 0.2 meters. Since the center of the optical beam coincides with the center of the FPA, i.e. $(x_0, y_0) = (0, 0),$ the probability of error performance for the case of 1 and 4 detectors is the same. However, Fig.~\ref{fig4} shows the difference in the performance related to $M=1$ and $M=4$ cases concerning the average probability of error when the beam position is not fixed. This is because the $M=4$ scheme weights the signal photons better than the noise photons when $(x_0, y_0) \neq (0, 0),$ and therefore, the average probability of error performance for $M=4$ system is better. In comparison, the probability of error performance for $M=1$ case is unaltered since the probability of error for this system is independent of $(x_0, y_0).$

\begin{figure}
	\centering
	\includegraphics[scale=1.5]{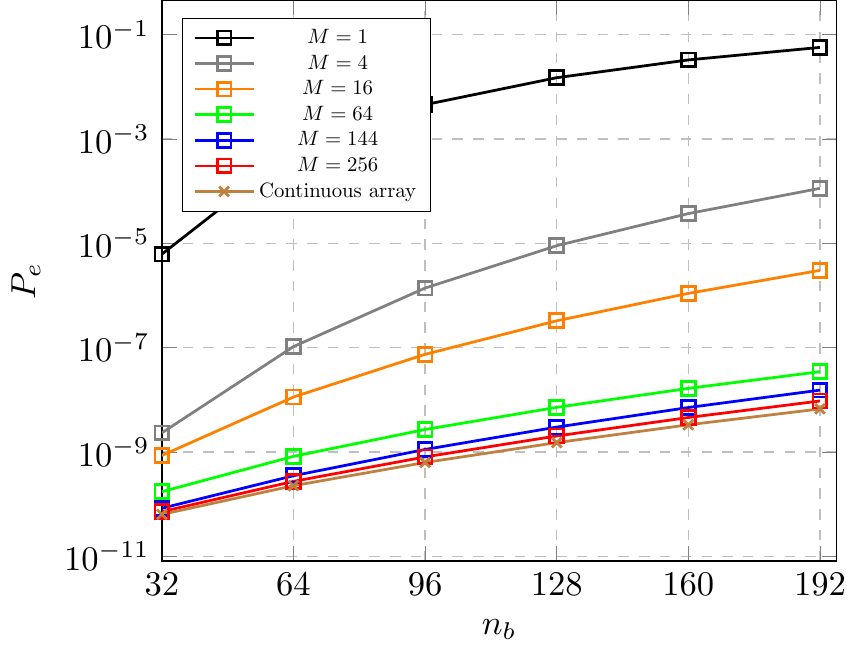}
	\caption{Probability of symbol error as a function of average number of noise photons for $I_0\rho^2(z)=800$, $\rho(z) = 0.1$ and $(x_0, y_0) = (0.4, 0.4)$. } \label{fig3}
\end{figure}

Fig.~\ref{fig3} describes the probability of error performance for the same signal power as before ($n_s = 50$ photons), but now the beam radius $\rho(z)$ has been reduced to 0.1 meters. In other words, we are using a narrower beam of the same power, and we note that the probability of error performance is better in this case as compared to the $\rho(z)=0.2$ meters scenario.

\begin{figure}
	\centering
	\includegraphics[scale=1.5]{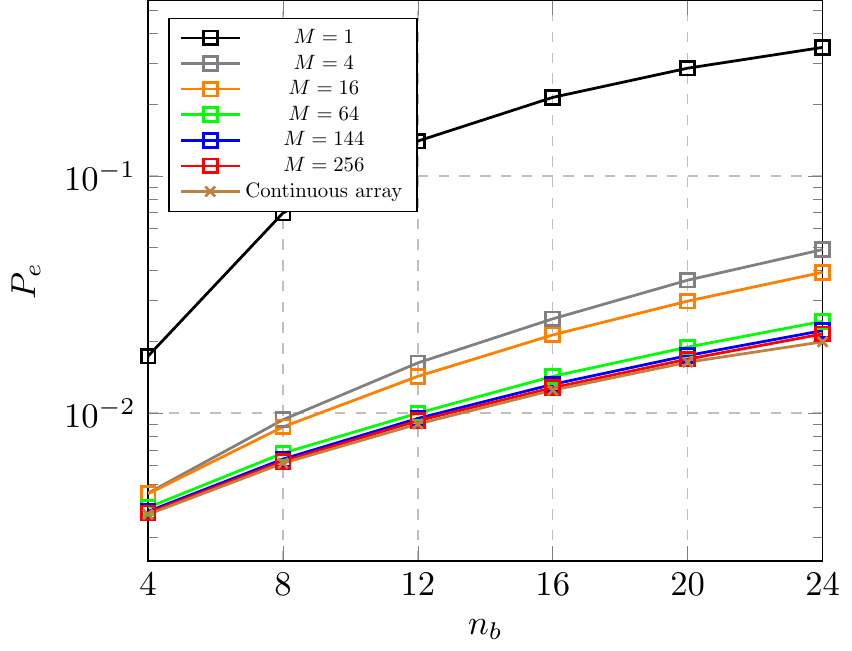}
	\caption{Probability of symbol error as a function of average number of noise photons for $I_0/\rho^2(z)=50$, $\rho(z) = 0.2$ and $(x_0, y_0) = (0.4, 0.4)$. } \label{fig5}
\end{figure}

\begin{figure}
	\centering
	\includegraphics[scale=1.5]{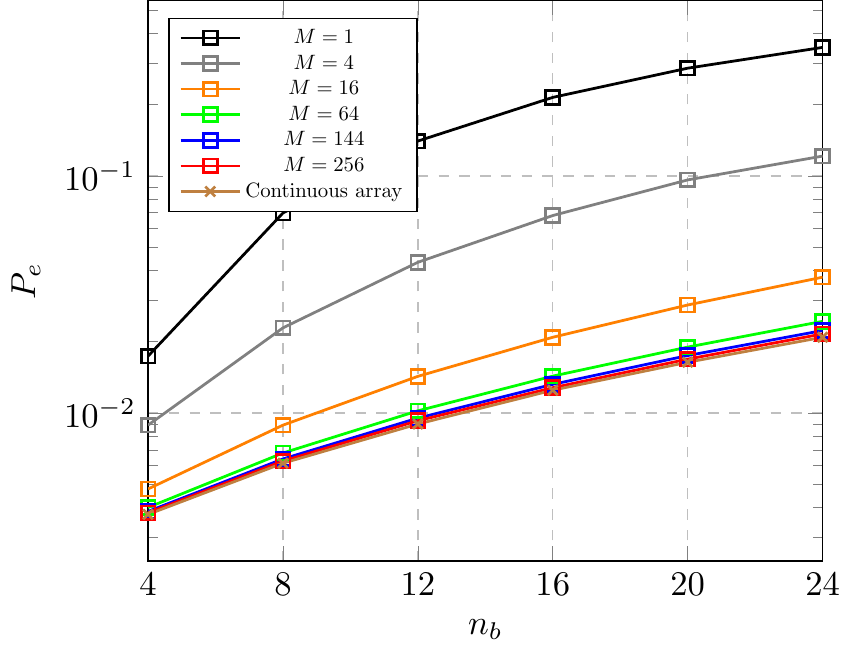}
	\caption{Probability of symbol error as a function of average number of noise photons for $I_0/\rho^2(z)=50$, $\rho(z) = 0.2$ and $(x_0, y_0)$ randomly sampled. } \label{fig6}
\end{figure}

Fig.~\ref{fig5} and Fig.~\ref{fig6} present the probability of error performance for low photon rate systems for which both $n_s$ and $n_b$ are smaller ($n_s = 12$  photons for the PPM slot). In terrestrial FSO, such low photon rate systems arise for foggy channels. Moreover, the photon rates for underwater optical communications can also be very small. 

All the plots are bounded from below by the probability of error performance with a continuous array. As we can see, the performance of the $12 \times 12$ and $16 \times 16$ arrays is quite close to the lower bound.

\subsection{Probability of error with imperfect knowledge of parameters} \label{uncertain}

In this section, we analyze the probability of error when there is uncertainty in the Gaussian beam parameters. These parameters include the signal-to-noise ratio factor $\frac{I_0}{\lambda_n},$ the beam radius $\rho(z),$ and the beam center position $(x_0, y_0).$  However, it is important to point out that the maximum likelihood detection does not require the estimation of the beam parameters for the single detector receiver. This is true because for $M=1$ case, there is only one term in the sum in \eqref{alphas}, and the factor $\alpha_m$ cancels out. Hence, the probability of error performance for a single detector array $(M=1)$ is independent with respect to the estimated values of beam parameters. 
\begin{figure}
	\centering
	\includegraphics[scale=1.5]{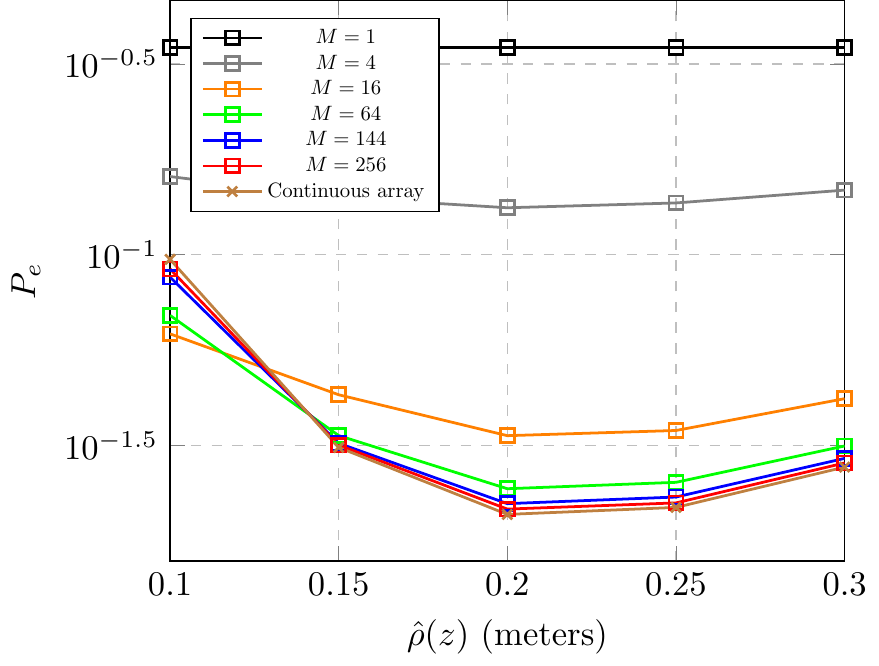}
	\caption{Probability of symbol error as a function of $\hat{\rho}(z)$ for $I_0/\rho^2(z)=50$, $\rho(z) = 0.2,$ $n_b=24$ and $(x_0, y_0)=(0.2, 0.2)$ . } \label{fig8}
\end{figure}

Fig.~\ref{fig8} shows the probability of error as a function of the estimate of $\rho(z)$ assuming that other beam parameter values are perfectly known. It can be seen that the probability of error is minimized when $\hat{\rho}(z) = \rho(z).$ Furthermore, the probability of error performance for  higher order detector arrays $(M \geq 16 )$  is a little higher when $\hat{\rho}(z) << \rho(z).$ Hence, we need to estimate $\rho(z)$ more accurately for higher order detector arrays. 
\begin{figure}
	\centering
	\includegraphics[scale=1.5]{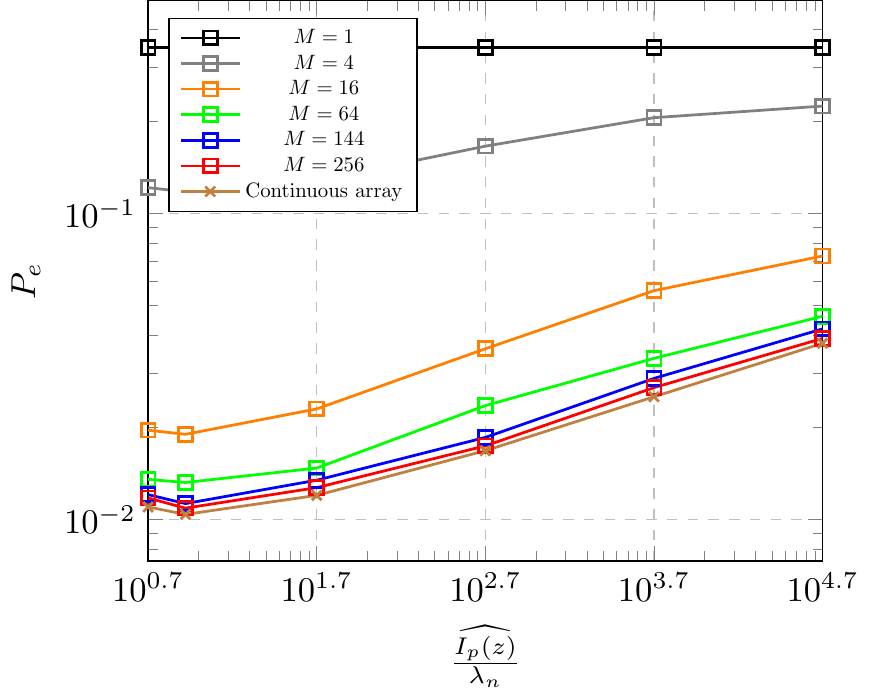}
	\caption{Probability of symbol error as a function of $\displaystyle \widehat{\frac{I_p(z)}{ \lambda_n}}$ for $\displaystyle \frac{I_p(z)}{ \lambda_n}=\frac{50}{6}$, $\rho(z) = 0.2,$ $n_b=24$ and $(x_0, y_0)=(0.2, 0.2)$ . } \label{fig9}
\end{figure}

Fig.~\ref{fig9} shows the error probability performance as a function of the estimated value of the signal-to-noise factor $I_p(z)/\lambda_n$ where $I_p(z) \triangleq I_0/\rho^2(z).$ From the plots, it can be seen that the probability of error performance is not affected significantly unless the estimation error is very large. Therefore, we can employ low complexity estimators (that do not have to be very precise) in order to estimate the signal-to-noise ratio.

\begin{figure}
	\centering
	\includegraphics[scale=1.5]{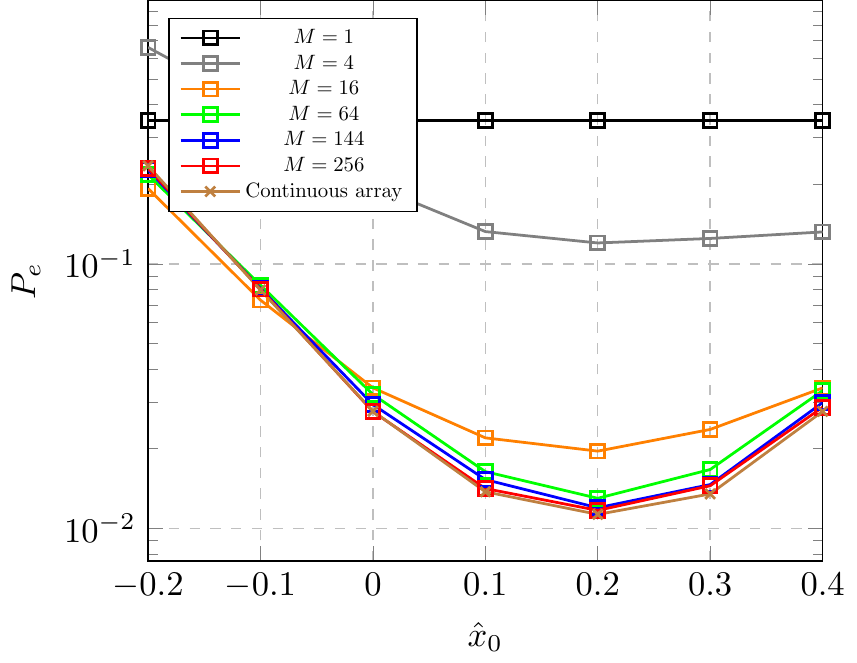}
	\caption{Probability of symbol error as a function of $\hat{x}_0$ for $\displaystyle \frac{I_p(z)}{ \lambda_n}=\frac{50}{6}$, $\rho(z) = 0.2,$ $n_b=24$ and $(x_0, y_0)=(0.2, 0.2)$ . } \label{fig10}
\end{figure}

In Fig.~\ref{fig10}, we analyze the probability of error performance as a function of $\hat{x}_0$ assuming that $y_0$ can be estimated exactly ($\hat{y}_0 = y_0$). Because the beam is circularly symmetric, the performance as a function of $\hat{y}_0$ will be the same if $\hat{x}_0 = x_0.$ Hence, we only consider the one dimensional case for the sake of simplicity. As can be seen from the plots in Fig.~\ref{fig10}, the probability of error is minimized for $\hat{x}_0 = x_0 =0.2,$ and the effect of the poor beam center position estimation on the error probability can be significant for higher order detector arrays. Hence, it is essential to estimate the beam position $(x_0, y_0)$ as accurately as possible in order to minimize the probability of error when $M$ is large.

\section{A brief complexity analysis}
In this section, we analyze the complexity overhead incurred with communications using detector arrays. In this regard, we consider the hard-decision maximum likelihood detector operating on a symbol-by-symbol basis for a $K-$PPM scheme. The detection algorithm will first compute a total of $K$ sums, where the $k$th sum is  $\sum_{m=1}^M \alpha_m Z_m^{(k)}$, and where $Z_m^{(k)}$ is the photon count in the $m$th detector during the $k$th slot of the PPM symbol for $k=1$ through $K$(see \eqref{alphas}). Thereafter, it will select the symbol $j$ if the $j$th sum has the maximum value. We can see that in order to compute a given $k$th sum, we need approximately $M$ real multiplies and $M$ real additions. Hence, the total complexity for a given symbol with such a receiver is $KM$ real additions and $KM$ real multiplies.

In addition to this, communications with detector arrays incur addition complexity overhead in terms of estimation of beam parameters as mentioned in Section~\ref{uncertain}. Let us assume that we use a naive estimator in order to ascertain the signal-to-noise ratio factor $I_0/\lambda_n$ since the estimation error in this parameter does not affect the probability of error significantly. However, that leaves us with the estimation of the beam radius $\rho(z)$ and the beam center position $(x_0, y_0).$ The complexity of different beam position estimators and trackers have been discussed in detail in \cite{Bashir1} and \cite{Bashir2} where the proposed estimators range from a naive centroid estimator to more sophisticated estimators like maximum likelihood estimator and particle filters. The general conclusion derived from these papers is that for any estimator, the computational complexity increases monotonically with $M.$ For example, with a centroid estimator, the complexity is roughly $2MK$ real additions and $2MK$ real multiplications. 

In order to estimate $\rho(z),$ a naive estimator can be proposed that will estimate the signal intensity in each cell of the detector array, and based on those estimates determine the estimated value of $\rho(z).$ For this naive estimator, the complexity of estimation goes up linearly with $M$ as well.

Let us denote the complexity of the beam position estimation by $C_{x_0, y_0}(M)$ and that of $\rho(z)$ by $C_{\rho(z)}(M).$ Then the total complexity for a symbol detection using an array of $M$ detectors is approximately $MK \text{ real multiplies} + MK \text{ real additions} + C_{x_0, y_0}(M) + C_{\rho(z)}(m)$.
\section{Conclusion} \label{conclude}
In this paper, we investigated the probability of error performance of a detector array or an FPA receiver with Gaussian beams in free-space optical communications. We found out that for the fixed signal-to-noise ratio and receiver size, the probability of error can be significantly minimized with an array of smaller detectors instead of using a single large detector, especially for poor signal-to-noise scenario. Moreover, the larger the number of detectors or cells $M$ in the array, the smaller the resulting probability of error. However, due to the law of diminishing returns, the improvement in the probability of error becomes smaller as the number of cells go from  $M$ to a higher number  for a large $M$. Having said that, detector arrays are also more helpful during the beam acquisition and tracking process, and the larger the number of cells, the better the tracking accuracy \cite{Bashir1}.

Finally, symbol detection with detector arrays result in a higher computational complexity that increases monotonically with $M$. So a trade-off has to be reached with respect to the probability of error performance and the complexity of the system.

\appendix
For continuous arrays, let there be $N$ photodetections during the slot interval (observation interval) $T_s$ where $N$ is a random variable.  We assume that the number and locations of the photodetections are modeled as a nonhomogeneous Poisson point process whose intensity function is given by \cite{Streit}
\begin{align}
\lambda(x,y,z) = \frac{I_0}{\rho^2(z)} \exp\left( \frac{-(x-x_0)^2 -(y-y_0)^2}{2 \rho^2(z)}  \right) + \lambda_n,
\end{align}
for the signal plus noise slot, and $\lambda_n$ for the noise only slot. 
  Let us denote the location of the $i$th photodetection on the two dimensional array be $\mathbf{x}_i \triangleq \begin{bmatrix}
x_i & y_i
\end{bmatrix}^T$. The loglikelihood ratio for the detection of the signal pulse during a given PPM slot is given by
\begin{align}
\ln \mathcal{L}(\mathbf{x}_1, \mathbf{x}_2, \dots, \mathbf{x}_n) = \ln \frac{p_1(\mathbf{x}_1, \mathbf{x}_2, \dots, \mathbf{x}_n)}{p_0(\mathbf{x}_1, \mathbf{x}_2, \dots, \mathbf{x}_n)}, \label{likeli}
\end{align}
where $p_1(\cdot)$ corresponds to the hypothesis $H_1$ that there is a signal pulse present in a given slot, $p_0(\cdot)$ corresponds to the hypothesis $H_0$ that there is no signal pulse present and $n$ is a realization of $N$.  The decision rule is 
\begin{align}
\ln \mathcal{L}(\mathbf{x}_1, \mathbf{x}_2, \dots, \mathbf{x}_n) \underset{H_0}{\overset{H_1}{\gtrless}} 0. \label{likelihood}
\end{align}

The likelihood functions can be expanded as
\begin{align}
&p_1(\mathbf{x}_1, \mathbf{x}_2, \dots, \mathbf{x}_n) = p_1(\mathbf{x}_1, \mathbf{x}_2, \dots, \mathbf{x}_N| N=n) P(\{ N = n\}) \nonumber \\
&= \prod_{i=1}^{n} \frac{1}{K_0}\left( \frac{I_0}{ \rho^2(z)} e^{ \frac{-(x_i-x_0)^2 - (y_i-y_0)^2}{2\rho^2(z)} } + \lambda_n  \right)  \nonumber \\
& \times \frac{\exp\left(-\int_{-1}^1 \int_{-1}^{1} \left(\frac{I_0}{ \rho^2(z)} e^{ \frac{-(x-x_0)^2 - (y-y_0)^2}{2\rho^2(z)} } + \lambda_n\right) \, dx\, dy \right)  \left[\int_{-1}^1 \int_{-1}^{1} \left(\frac{I_0}{ \rho^2(z)} e^{ \frac{-(x-x_0)^2 - (y-y_0)^2}{2\rho^2(z)} } + \lambda_n\right) \, dx\, dy\right]^n } { n!} \label{likely}
\end{align}
where we have assumed that the detector array extends from -1 to 1 in each dimension in the two dimensional space, and the random variables $\mathbf{x}_0, \mathbf{x}_1, \dots, \mathbf{x}_n$ are conditionally independent. The quantity $K_0$ is a normalization constant. Taking the natural logarithm of both sides of \eqref{likely}, we have that
\begin{align}
&\ln p_1(\mathbf{x}_1, \mathbf{x}_2, \dots, \mathbf{x}_n)=  -\int_{-1}^1 \int_{-1}^{1} \left(\frac{I_0}{ \rho^2(z)} e^{ \frac{-(x-x_0)^2 - (y-y_0)^2}{2\rho^2(z)} } + \lambda_n\right) \, dx\, dy \nonumber \\
&+ n \ln \left(\int_{-1}^1 \int_{-1}^{1} \left(\frac{I_0}{ \rho^2(z)} e^{ \frac{-(x-x_0)^2 - (y-y_0)^2}{2\rho^2(z)} } + \lambda_n\right) \, dx\, dy \right) - \ln n! + \sum_{i=1}^{n} \ln \left[ \frac{1}{K_0}\left( \frac{I_0}{ \rho^2(z)} e^{ \frac{-(x_i-x_0)^2 - (y_i-y_0)^2}{2\rho^2(z)} } + \lambda_n \right) \right]. \label{likeli2}
\end{align}
For the noise only hypothesis, $p_0(\mathbf{x}_1, \dots, \mathbf{x}_n | N=n ) = (\lambda_n A_d)^n$, where $A_d$ is the total area of the detector array. Hence, by following the similar arguments as used in the derivation of \eqref{likeli2}, it can be easily shown that 
\begin{align}
\ln p_0(\mathbf{x}_1, \mathbf{x}_2, \dots, \mathbf{x}_n) = 2n \ln (\lambda_n A_d) - \lambda_n A_d - \ln n!.
\end{align}
Finally, the decision rule in \eqref{likelihood}, after a few manipulations, can be rewritten as 
\begin{align}
& n \ln \left(\int_{-1}^1 \int_{-1}^{1} \left(\frac{I_0}{ \rho^2(z)} e^{ \frac{-(x-x_0)^2 - (y-y_0)^2}{2\rho^2(z)} } + \lambda_n\right) \, dx\, dy \right)+ \sum_{i=1}^{n} \ln \left[ \frac{1}{K_0}\left( \frac{I_0}{ \rho^2(z)} e^{ \frac{-(x_i-x_0)^2 - (y_i-y_0)^2}{2\rho^2(z)} } + \lambda_n \right) \right] \nonumber \\
&- 2n \ln (\lambda_n A_d)      \underset{H_0}{\overset{H_1}{\gtrless}}    \int_{-1}^1 \int_{-1}^{1} \left(\frac{I_0}{ \rho^2(z)} e^{ \frac{-(x-x_0)^2 - (y-y_0)^2}{2\rho^2(z)} } \right) \, dx\, dy, 
\end{align}
where the quantity on the right hand side is just a constant (signal power).

Let us assume that a symbol $j$ was transmitted in a $K-$PPM scheme. A hard-decision maximum likelihood receiver will decide that a symbol $j, 1 \leq j \leq K$ was transmitted if \cite{Bashir3}
\begin{align}
\ln \mathcal{L}(\mathbf{x}^{(j)}_1, \mathbf{x}^{(j)}_2, \dots, \mathbf{x}^{(j)}_{n_j}) > \ln \mathcal{L}(\mathbf{x}^{(k)}_1, \mathbf{x}^{(k)}_2, \dots, \mathbf{x}^{(k)}_{n_k}), \; \text{ for every }k, \; 1 \leq k \leq K, \; k \neq j.
\end{align}
where $\mathbf{x}^{(j)}_i$ and $\mathbf{x}^{(k)}_i$ is the location of  $i$th photodetection on the continuous array in the $j$th and $k$th slots of the PPM symbol, respectively. Moreover,  $n_j$ and $n_k$ are the photodetection counts in the $j$th and $k$th slots, respectively, and the function $\mathcal{L}(\cdot)$ is defined in \eqref{likeli}. It should be noted that $\mathbf{x}_i^{(j)}$ is the $i$th photodetection location corresponding to the signal plus noise slot, and $\mathbf{x}_i^{(k)}$ corresponds to a noise only slot.  The probability of the correct decision at the receiver, given that a symbol $j$ was transmitted, is given by 
\begin{align}
P_{c|j} = P\left(   \left\{  \    \ln \mathcal{L}(\mathbf{x}^{(j)}_1, \mathbf{x}^{(j)}_2, \dots, \mathbf{x}^{(j)}_N) - \ln \mathcal{L}(\mathbf{x}^{(k)}_1, \mathbf{x}^{(k)}_2, \dots, \mathbf{x}^{(k)}_N) > 0    \right\}   \right)^{K-1}, \label{Pc}
\end{align}
where we have used the fact that $\mathbf{x}_i^{(j)}$ and $\mathbf{x}_l^{(k)}$ are independent random variables whenever $i \neq l$ or $j \neq k$. Substituting the expression of the loglikelihood in \eqref{Pc}, the final expression for the probability of a correct decision yields the form
\begin{small}
\begin{align}
P_{c|j} &= \left( \left\{  (n_j - n_k) \ln \left(\int_{-1}^1 \int_{-1}^{1} \left(\frac{I_0}{ \rho^2(z)} e^{ \frac{-(x-x_0)^2 - (y-y_0)^2}{2\rho^2(z)} } + \lambda_n\right) \, dx\, dy \right)+ \sum_{i=1}^{n_j} \ln \left[ \frac{1}{K_0}\left( \frac{I_0}{ \rho^2(z)} e^{ \frac{-(x^{(j)}_i-x_0)^2 - (y^{(j)}_i-y_0)^2}{2\rho^2(z)} } + \lambda_n \right) \right] \right. \right.  \nonumber \\
& \left. \left. -  \sum_{i=1}^{n_k} \ln \left[ \frac{1}{K_0}\left( \frac{I_0}{ \rho^2(z)} e^{ \frac{-(x^{(k)}_i-x_0)^2 - (y^{(k)}_i-y_0)^2}{2\rho^2(z)} } + \lambda_n \right) \right]   - 2(n_j - n_k) \ln (\lambda_n A_d)      > 0  \right\} \right)^{K-1},
\end{align}
\end{small}
where $\mathbf{x}_i^{(j)} = \begin{bmatrix}
x_i^{(j)} & y_i^{(j)}
\end{bmatrix}^T$. For an equiprobable symbol transmission, the probability of error is,  $P_e = 1 - P_{c} = 1 - P_{c|j}.$
\ifCLASSOPTIONcaptionsoff
  \newpage
\fi


%





\begin{thebibliography}{1}

\bibitem{Kaymak}
 Y.~Kaymak, R.~Rogas-Cessa, J.~Feng, N.~Ansari, M.~Zhou and T.~Zhang, ``A survey on acquisition, tracking and pointing mechanisms for mobile free-space optical communications,'' \emph{IEEE Communication Surveys and Tutorials,}, vol.~20, no.~2, 2018.
 
 \bibitem{Srinivasan1}
 M.~Srinivasan, K.~S.~Andrews, W.~H.~Farr, A.~Wong,
 ``Photon counting detector array algorithms for deep space optical
 communications,'' \emph{Proc. SPIE 9739, Free-Space Laser Communication and
 	Atmospheric Propagation XXVIII, 97390X}, doi:
 10.1117/12.2217971, 2016.
 
 
 
 \bibitem{Vilnrotter}
 V.~Vilnrotter and M.~Srinivasan, ``Adaptive detector arrays for optical communications receivers,'' \emph{TMO Progress Report 42-141,} 2000.
 
 \bibitem{Srinivasan}
 M.~Srinivasan and V.~Vilnrotter, ``Avalanche photodiode arrays for optical communications receivers,'' \emph{TMO Progress Report 42-144,} 2001.
 
 \bibitem{Vilnrotter1}
 V.~Vilnrotter, C.~W.~Lau, M.~Srinivasan, R.~Mukai and K.~Andrews, ``An optical array receiver for deep space communication through atmospheric turbulence,'' \emph{IPN Progress Report 42-154,} 2005.
 
 \bibitem{Kaushal}
 H.~Kaushal, V.~K.~Jain and S.~Kar, \emph{Free Space Optical Communications,} Springer (India) Pvt. Ltd. 2017.
 

 
 \bibitem{Goodman}
 J.~W.~Goodman, \emph{An Introduction to Fourier Optics}, Roberts and Company Publishers, Eaglewood CO, 2005.
 
  \bibitem{Srinivasan2}
 M.~Srinivasan, V.~Vilnrotter, M.~Troy and K.~Wilson, ``Adaptive optics communication performance analysis,'' \emph{IPN-Progress Report 42-158}, 2004.
 
 
 
 \bibitem{Najafi}
 M.~Najafi, H.~Ajam, V.~Jamali, P.~D.~Diamantoulakis, G.~K.~Karagiannidis and R.~Schober, ``Statistical modeling of FSO fronthaul channel for drone-based networks,'' \emph{arXiv:1711.00120v1}, October 2017.
 
 
 \bibitem{Bashir1} 	M.~S.~Bashir and M.~R.~Bell, {``Optical beam position estimation in free-space optical communication,''} \emph{IEEE Transactions on Aerospace and Electronic Systems}, Vol.~52, No.~6, December 2016.
 
 
 
 \bibitem{SnyderandMiller} 
 D.~L.~Snyder, M.~I.~Miller, \emph{Random Point Processes in Time and Space}, New York, NY: Springer-Verlag, 1991.
 
 
 
 \bibitem{Bashir2} M.~S.~Bashir and M.~R.~Bell, {``Optical beam position tracking in free-space optical communication systems,''} \emph{IEEE Transactions on Aerospace and Electronic Systems}, Vol.~20, No.~2, April 2018.
 
 \bibitem{Bashir3}
  M.~S.~Bashir and M.~R.~Bell, {``The impact of optical beam position estimation on the probability of error in free-space optical communications,''} \emph{IEEE Transactions on Aerospace and Electronic Systems}, accepted for publication, June 2018.
  
  \bibitem{Peyronel}
  T.~Peyronel , K.~J.~Quirk , S.~C.~Wang and T.~G.~Tiecke, ``Luminescent detector for free-space optical
  communication,'' \emph{Optica}, Vol.~3, No.~7, 2016.
  
  
  
 
\bibitem{Hajek}
B.~Hajek, \emph{Random Processes for Engineers,} Cambridge University Press, 2015. 

\bibitem{Gallager}
~R.~G.~Gallager, \emph{Information Theory and Reliable Communication}, New York, NY: John Wiley and Sons Inc., 1968, pp. 127--129. 

\bibitem{Streit}
R.~L.~Streit, \emph{Poisson Point Processes: Imaging, Tracking and Sensing,} Springer Science+Business Media, LLC, 233 Spring Street, New York, NY, 2010.
\end{thebibliography}
\end{document}